\documentclass[aps,prl,showpacs,showkeys,floatfix,nofootinbib,twocolumn,
10pt]{revtex4-1}
\usepackage{natbib}
\usepackage{ulem}
\usepackage{color}
\usepackage{graphicx}
\usepackage{epsfig}
\usepackage{bm}
\usepackage{amsthm}
\usepackage{amsfonts}
\usepackage{float}
\usepackage{amsmath,amssymb}
\usepackage{graphicx}
\usepackage[caption = false]{subfig}
\usepackage{hyperref}
\usepackage{cleveref}

\begin{document}
\title{
Novel scheme for parametrizing the chemical freeze-out surface
in Heavy Ion Collision Experiments}
\author{Sumana Bhattacharyya}
\email{response2sumana@jcbose.ac.in}
\author{Deeptak Biswas}
\email{deeptak@jcbose.ac.in}
\author{Sanjay K. Ghosh}
\email{sanjay@jcbose.ac.in}
\author{Rajarshi Ray}
\email{rajarshi@jcbose.ac.in}
\author{Pracheta Singha}
\email{pracheta@jcbose.ac.in}
\affiliation{
Department of Physics,
\\ \& \\ 
	Center for Astroparticle Physics \& Space Science,\\
Bose Institute, EN-80, Sector-5, Bidhan Nagar, Kolkata-700091, India 
}
%
%
\begin{abstract}
We introduce a new prescription for obtaining the chemical freeze-out
parameters in the heavy-ion collision experiments using the Hadron
Resonance Gas model. The scheme is found to reliably estimate the
freeze-out parameters and predict the hadron yield ratios, which
themselves were never used in the parametrization procedure. 

\end{abstract}

%
\maketitle
%
{\it Introduction:} $-$
Strongly interacting matter is expected to exhibit a rich phase
structure under extreme conditions of temperature and density.  Exotic
phases with quasi-free quarks and gluons may have existed at high
temperatures in the very early universe~\cite{Kolb:1990vq}. Even today,
in the core of compact stars with high baryon densities, various exotic
phases like color superconductivity, color superfluidity, may be
present~\cite{doi:10.1142/9789812810458_0043}.

Direct signatures of these phases can only be accessed in experiments
with relativistic nuclear collisions that are being pursued at CERN
(France/Switzerland) and BNL (USA), and also to be carried out at GSI
(Germany) and JINR (Russia). In the canonical picture of heavy-ion
collision (HIC), the high density fireball formed, is expected to
thermalize rapidly, expand out fast and then cool down quickly. As the
system expands, the inter-particle distances increase, and subsequently
all thermal and chemical interactions freeze-out. Finally the detected
strongly interacting particles are the hadrons and their resonances
which may be in chemical equilibrium~\cite{Hagedorn:1965st,
Hagedorn:1980kb}. Later, it has been argued that a transient
partonic phase is more likely to drive the onset of equilibration
in the system~\cite{PhysRevD.22.2793,PhysRevLett.48.1066,Koch:1986ud}.

In a pioneering work~\cite{PhysRevLett.81.5284}  using the Hadron
Resonance Gas (HRG) model, it was argued that the freeze-out surface may
be universally characterized by the average energy per hadron to have a
value of 1 GeV. Subsequently there has been a huge interest in studying
the properties of strongly interacting matter using HRG
model~\cite{Rischke:1991ke, Cleymans:1992jz, BraunMunzinger:1994xr,
Cleymans:1996cd, Yen:1997rv, Heinz:1998st, BraunMunzinger:1999qy,
Cleymans:1999st, BraunMunzinger:2001ip, Becattini:2003wp,
BraunMunzinger:2003zd, Karsch:2003zq, Tawfik:2004sw, Becattini:2005xt,
Andronic:2005yp, Andronic:2008gu, Manninen:2008mg, Tiwari:2011km,
Begun:2012rf, Andronic:2012ut, Fu:2013gga, Tawfik:2013eua, Garg:2013ata,
Bhattacharyya:2013oya, Albright:2014gva, Kadam:2015xsa, Kadam:2015fza,
Kadam:2015dda, Albright:2015uua, Bhattacharyya:2015zka,
Bhattacharyya:2015pra, Begun:2016cva, Bhattacharyya:2017gwt,
Andronic:2017pug, Dash:2018can}. This model has successfully described
hadron yields from AGS to LHC energies~\cite{BraunMunzinger:1994xr,
Cleymans:1996cd, Yen:1997rv, BraunMunzinger:1999qy, Cleymans:1999st,
BraunMunzinger:2001ip, BraunMunzinger:2003zd, Becattini:2003wp,
Cleymans:2004hj, Letessier:2005qe, Becattini:2005xt, Cleymans:2005xv, 
Andronic:2005yp, Manninen:2008mg, Andronic:2008gu, Tiwari:2011km, 
Chatterjee:2013yga, Alba:2014eba, Chatterjee:2015fua, Adak:2016jtk}. 
In this context the discussions of multi-strange 
enhancement~\cite{PhysRevLett.43.1292} and
saturation of strangeness~\cite{PhysRevLett.48.1066} in a quark-gluon
phase came up. Some authors also considered possible under-saturation of
strangeness in the observed spectrum~\cite{Koch:1986ud},
\cite{Becattini:2003wp}.  Bulk properties of hadronic matter have also
been studied in this model~\cite{Karsch:2003zq, Tawfik:2004sw,
Andronic:2012ut}. Moreover, role of various undiscovered resonance
states in determining the freeze-out surface has been
investigated~\cite{Chatterjee:2017yhp}. There has been recent attempts
to address the freeze-out conditions even from the first principle
lattice QCD \cite{Gupta1525, Bazavov:2012vg}. The general perception
from all these studies is that at freeze-out the hadrons are in
thermodynamic equilibrium.

Here we propose a novel fitting procedure for extraction of thermal 
parameters from experimental hadron yields with the ideal HRG model 
and show that the ratios of the yields are very well reproduced for 
experiments over a wide range of collision energies. 

~\\
{\it HRG Model:} $-$
The grand canonical partition function of the hadron resonance gas is
given by,

\begin {equation}
\ln Z^{ideal}=\sum_i \ln Z_i^{ideal},
\end{equation}
The sum runs over all hadrons and resonances.  The thermodynamic
potential for $i$'th species is given as,
\begin{equation}
\ln Z_i^{ideal}=\pm \frac{Vg_i}{(2\pi)^3}\int d^3p
\ln[1\pm\exp(-(E_i-\mu_i)/T)],
\end{equation}
where the upper sign is for baryons and lower for mesons. Here $V$ is
the volume, $T$ is the temperature, and for the $i^{th}$ species of
hadron, $g_i$, $E_i$ and $m_i$ are respectively the degeneracy factor,
energy and mass, while $\mu_i=B_i\mu_B+Q_i\mu_Q+S_i\mu_S$ is the
chemical potential, with $B_i$, $Q_i$ and $S_i$ denoting the baryon
number, electric charge and strangeness respectively. Here $\mu_B$,
$\mu_Q$ and $\mu_S$ are the baryon, electric and strangeness chemical
potentials respectively. For a thermalized system the number density
$n_i$ can be calculated from partition function, which is given as,

\begin{equation}
n_i(T,\mu_B,\mu_Q,\mu_S)=\frac{g_i}{{(2\pi)}^3} \int\frac{d^3p}
{\exp[(E_i-\mu_i)/T]\pm1}.
\end{equation}

The thermal parameters $(T,\mu_B,\mu_Q,\mu_S)$ may then be obtained by
fitting experimental hadron yields to the model parametrization via the
relation between the rapidity density for $i$'th detected hadron to the
corresponding number density in the HRG model~\cite{Manninen:2008mg},
\begin{equation}
\frac{dN_i}{dy}|_{Det}={\frac{dV}{dy}}n_i^{Tot}|_{Det}
\end{equation}
where the subscript $Det$ denotes the detected hadrons. Here,
\begin{eqnarray}
&n_i^{Tot}& ~=~ n_i(T,\mu_B,\mu_Q,\mu_S)~ + 
\nonumber \\
&\sum_j& n_j(T,\mu_B,\mu_Q,\mu_S) \times Branch~Ratio (j
\rightarrow i)
\end{eqnarray}
where the summation is over the heavier resonances $j$ that decay to the
$i^{th}$ hadron. Usually the systematics due to the volume factor is
removed by considering hadron yield ratios. Thereafter one needs four
equations to solve for the four freeze-out parameters $T$, $\mu_B$,
$\mu_Q$ and $\mu_S$. The $\mu_Q$ and $\mu_S$ are fixed by imposing
the constraints~\cite{Alba:2014eba},

\begin{equation}
\frac{\sum_i n_i (T, \mu_B, \mu_Q, \mu_S) Q_i}{\sum_i n_i (T,
\mu_B, \mu_Q, \mu_S) B_i}=r
\label{eq.nbq}
\end{equation}
and
\begin{equation}
\sum_i n_i (T, \mu_B, \mu_S, \mu_Q) S_i=0
\label{eq.ns}
\end{equation}
\noindent
where $r$ is net-charge to net-baryon number ratio of the colliding
nuclei. For example, in Au + Au collisions $r = N_p /(N_p + N_n)=0.4$,
with $N_p$ and $N_n$ denoting the number of protons and neutrons in the
colliding nuclei. Thereafter the $T$ and $\mu_B$ are conventionally
fitted by optimizing the $\chi^2$ of the multiplicity ratios with
respect to $T$ and $\mu_B$. Here the $\chi^2$ is defined as,
\begin{eqnarray}
\chi^2=\sum_i \frac{(Ratio^{Model}_{i}-Ratio^{Expt}_{i})^2}
{{\sigma_i}^2}
\label{eq.chis}
\end{eqnarray}
and the two minimization equation correspond to,
\begin{equation}
\frac{\partial \chi^2}{\partial x}=0; ~~{\rm for}~~
x \subset \{T,\mu_B\}.
\label{eq.chism}
\end{equation}
Thus the freeze-out parameters are obtained from
Eq.~(\ref{eq.nbq}$-$\ref{eq.chism}). A satisfactory solution is obtained
if $\chi^2$ over degrees of freedom (dof) is close to 1
\cite{NRf77:1996}. 

~\\
{\it The New Approach:} $-$
The value of $\chi^2$ and extracted set of parameters strongly depend on
the set of multiplicity ratios chosen. Selecting a particular set of
ratios may bias the minimization process \cite{Andronic:2005yp}. This is
possibly due to the fact that the individual multiplicities or their
ratios are not independent quantities. If they were then one could
replace Eq.(\ref{eq.chis}$-$\ref{eq.chism}) by equating any two
multiplicity ratios from the model and experimental data.  These along
with Eq.~(\ref{eq.nbq}$-$\ref{eq.ns}) would then give the freeze-out
parameters. Usually if a solution is obtained in this method, the
prediction of other multiplicity ratios are found to be significantly
away from the experimental data. This is why a $\chi^2$ analysis is
done including as many independent individual multiplicity ratios as
available from experimental data. As a result the prediction of hadron
yields in the $\chi^2$ fit seems to be {\it built in by default}.
Nonetheless the $\chi^2$ fit does produce a set of freeze-out parameters
commensurate with all the multiplicity ratios to some extent.

Here we ask if it is possible to consider any truly independent
observable multiplicity ratios so that simply equating them between the
model and the experimental data one can obtain the freeze-out
parameters. And finally whether they predict all individual hadron
multiplicity ratios satisfactorily. In this direction we again note that
for strong interactions to be in chemical equilibrium, there are five
independent thermodynamic variables $-$ $V$, $T$, $\mu_B$, $\mu_Q$ and
$\mu_S$. Obviously the three net conserved charges are independent, but
are not enough to determine the five thermodynamic parameters. The three
corresponding total charges though not conserved are however
independent. We therefore perform our analysis of freeze-out data based
on this observation.

Considering the three net charges and three total charges we have one
more independent quantity than that required to determine five
thermodynamic variables. We can then assume a constrain like entropy
conservation. In terms of the observed hadrons this imposes a constrain
on the total number of particles \cite{Landau:1953gs}.  We can then uniquely
determine all the five thermodynamic variables of the system.  Following
the general practice we considered ratios of the different charges to
scale out the volume and any other systematics. We thus need to consider
four independent ratios. Fortunately Eq.~(\ref{eq.nbq}$-$\ref{eq.ns})
are already of the desired form (Eq.(\ref{eq.ns}) may be read as the
ratio of net strangeness to total strangeness to be zero). Here,
instead of the two optimization equations Eq.~\ref{eq.chism}, with
respect to $T$ and $\mu_B$ we introduce two new independent equations.
We choose the net baryon number normalized to the total baryon number
and the net baryon number normalized to the total hadron yield, to form
the other two equations, as given below.

\begin{eqnarray}
\frac{\sum_i^{Det} B_i \frac{dN_i}{dY}}{\sum_i^{Det} |B_i|
\frac{dN_i}{dY}}
&=& \frac{\sum_i^{Det} B_i n_i^{Tot}}{\sum_i^{Det} |B_i| n_i^{Tot}} 
\label{eq.conserveb} \\
\frac{\sum_i^{Det} B_i \frac{dN_i}{dY}}
{\sum_i^{Det}\frac{dN_i}{dY}}
&=& \frac{\sum_i^{Det} B_i
n_i^{Tot}}{\sum_i^{Det} n_i^{Tot}} 
\label{eq.t}
\end{eqnarray}
The ratios on the left hand side of the above equations consists of the
rapidity density of hadron yields measured in the HIC experiments and
those on the right are the number densities calculated in the HRG model.
The sum runs only over the identified hadrons for which the yield data
are available. The equations are clearly unique and independent of each
other if a sufficient number of identified hadrons are involved.

~\\
{\it Data Analysis:} $-$
We have used AGS~\cite{Ahle:1999uy, Ahle:2000wq, Klay:2003zf,
Klay:2001tf, Back:2001ai, Blume:2011sb, Back:2000ru, Barrette:1999ry, 
Back:2003rw}, SPS~\cite{Alt:2007aa, Alt:2005gr, Afanasiev:2002mx, 
Afanasev:2000uu, Bearden:2002ib, Anticic:2003ux, Antinori:2004ee, 
Antinori:2006ij, Alt:2008qm, Alt:2008iv, Anticic:2003ux}, 
RHIC~\cite{Kumar:2012fb, Das:2012yq, Adler:2002uv, Adams:2003fy,
Zhu:2012ph, Zhao:2014mva, Kumar:2014tca, Das:2014kja, Abelev:2008ab,
Aggarwal:2010ig, Abelev:2008aa, Adcox:2002au, Adams:2003fy,
Adler:2002xv, Adams:2006ke, Adams:2004ux, Kumar:2012np, Adams:2006ke}
and LHC~\cite{Abelev:2012wca, Abelev:2013xaa, ABELEV:2013zaa,
Abelev:2013vea} data for our analysis. STAR BES data has been used
following~\cite{Chatterjee:2015fua, Nasim:2015gua, Adamczyk:2017iwn}. 
In the present study we have only taken mid-rapidity data for the most 
central collisions.

In our HRG spectrum we have used all hadrons up to 2 GeV which are
confirmed with known degrees of freedom. The masses and branching ratios
used are as given in~\cite{Wheaton:2004qb, Tanabashi:2018oca}.  But data
are available for only a few hadrons at various collision energies.  The
identified hadrons used to obtain the freeze-out parameters are,
$\pi^\pm$ (139.57 MeV), $k^{\pm}$ (493.68 MeV), $p$,$\bar{p}$ (938.27
MeV), $\Lambda$,$\bar{\Lambda}$ (1115.68 MeV), $\Xi$,$\bar{\Xi}$
(1321.71 MeV).  All those hadrons reported at one collision energy may
not be available in another. For example, we could not find the
$\bar{\Lambda}$ yield at LHC. At this energy we assumed $\bar{\Lambda}$
yield to be same as that reported for ${\Lambda}$.  Similarly, we did
not use $\Omega$ data for any parametrization, as the individual yields
of $\Omega^+$ and $\Omega^-$ are not available for most of the
$\sqrt{s}$.  Also, $\phi$ (1019.46 MeV) has been excluded from the
fitting as it is already included in the model through its strong decay
channel to kaon.  For the lower AGS energies we could not find any
anti-baryon data. There we have used $\Lambda$ to proton ratio and total
baryon to total hadron yield as a substitution of
Eq.~{\ref{eq.conserveb}} and Eq.~{\ref{eq.t}}.

The equations
Eq.~(\ref{eq.nbq}$-$\ref{eq.ns}$-$\ref{eq.conserveb}$-$\ref{eq.t}), are
highly non-linear and are solved numerically using the Broyden's method
with a convergence criteria of $10^{-6}$ or better. We had to tune the
initial conditions accordingly for different $\sqrt{s}$ to achieve
desired convergence accuracy. The variances of the fitted parameters
were obtained by extracting the freeze-out parameters at the extremum
values of the hadron yields given by the experimental variances.

\begin{figure}[!htb]
\subfloat[]{
{\includegraphics[scale=0.7]{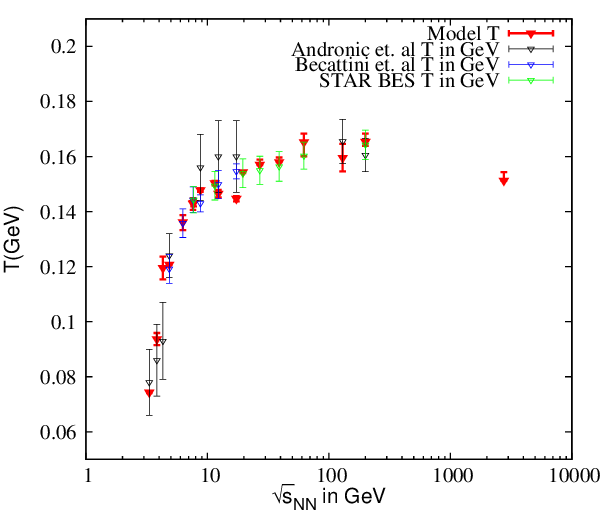}}
\label{fg.Ta}
}\\
\subfloat[]{
{\includegraphics[scale=0.7]{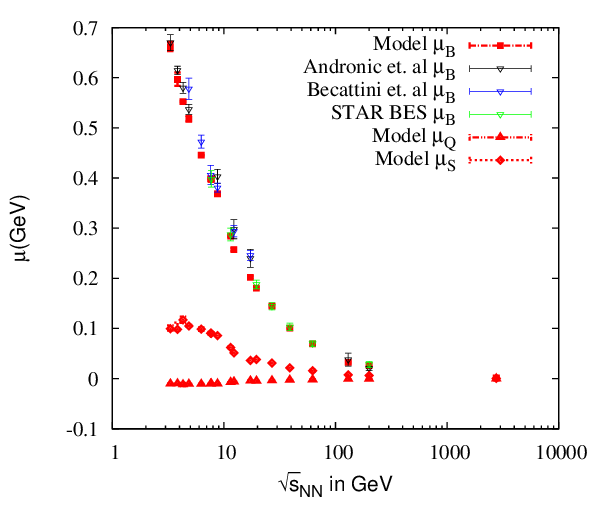}}
\label{fg.Tb}
}\\
\caption{
\label{fg.T}
Variation of $T$, $\mu_B$, $\mu_Q$, $\mu_S$ with $\sqrt{s}$}
\end{figure}

~\\
{\it Freeze-out Parameters:} $-$
The freeze-out parameters are depicted in Fig.~\ref{fg.T}. The general
behavior as well as the quantitative estimates are commensurate with
those in the existing literature. The variation of the freeze-out
temperature with the center of mass energy $\sqrt{s}$ is shown in
Fig.~\ref{fg.Ta}.  As expected, the temperature increases with
increasing $\sqrt{s}$, and approaches a saturation~\cite{Hagedorn:1965st},
except at the LHC energy where the temperature is lower. This is
probably due to the lower yield of protons at LHC~\cite{Abelev:2013vea}. 

In Fig.~\ref{fg.Tb} the various chemical potentials are shown as
functions of $\sqrt{s}$. The baryon chemical potential $\mu_B$ decreases with 
increasing $\sqrt{s}$, which is usually understood as follows. For low 
collision energies a significant amount of baryons may be deposited in the
collision region (baryon stopping). On the other hand at high collision
energies the colliding baryons may almost pass through each other and
get deposited outside the collision region. Similarly the electric
charge chemical potential $\mu_Q$ should have also followed the same
trend as the colliding nuclei only consisted of positively charged
protons. However $\mu_Q$ remains negative throughout, approaching zero
for increasing $\sqrt{s}$. Here, the neutrons in the colliding nuclei
(lead or gold), being more abundant than the protons, induce an isospin
dominance in favor of $\pi^-$ than $\pi^+$. The pions being the lightest
charged particles, dictates the sign of $\mu_Q$.  On the other hand,
strangeness production is expected to be dominant at higher baryon
densities due to the possible redistribution of Fermi momentum among
larger degrees of freedom lowering the Fermi
energy~\cite{Witten:1984rs}. Though it is not clear whether this picture
should hold in the HIC scenario, the fitted strangeness chemical
potential $\mu_S$, does indeed show such a behavior.

In Fig.\ref{fg.Ta} and Fig.\ref{fg.Tb} we have also shown comparison of
our results with those in literature. Specifically we have used results
from \cite{Andronic:2005yp, Becattini:2005xt} for SPS and RHIC energies.
For BES energy range we have compared our results with
\citep{Adamczyk:2017iwn}. The general agreement in almost all the
studies is apparant, though there are certain variations in the models.
For example Ref.~\citep{Adamczyk:2017iwn} considered a strangeness
suppression factor and have assumed $\mu_Q$ to be zero.  Similarly
Ref.~\cite{Becattini:2003wp, Becattini:2005xt, Cleymans:2005xv,
Manninen:2008mg, Andronic:2005yp} have considered a strangeness
suppression factor. The system volume was extracted as a parameter in
Ref.~\cite{Andronic:2005yp, Andronic:2008gu, Andronic:2012ut}.  In an
other work~\cite{Letessier:2005qe} a light quark fugacity factor has
also been introduced, which may have significant effect on the value of
extracted parameter set and $\chi^2/dof$. 

~\\
{\it Hadron Yield Ratios:} $-$
With the freeze-out parameters obtained, we now discuss the various
predicted hadron yield ratios. Though the hadronic yields were used in
the analysis, none of their individual ratios were part of the equations
solved, and are therefore quite independent predictions from the model.
The only exception is the use of the single ratio $\Lambda/p$ for the
lower AGS energies. From the experimental data the variations in the
yield ratios are obtained from those of the individual yields using
standard error propagation method~\cite{knoll2000radiation}. We have
considered both systematic and statistical errors and the total error
for a particular yield was obtained in quadrature. Here we discuss some
important representative hadron ratios.  The predictions of other hadron
yields also came out satisfactorily and will be presented elsewhere.

\begin{figure}[!htb]
{\includegraphics[scale=0.7]{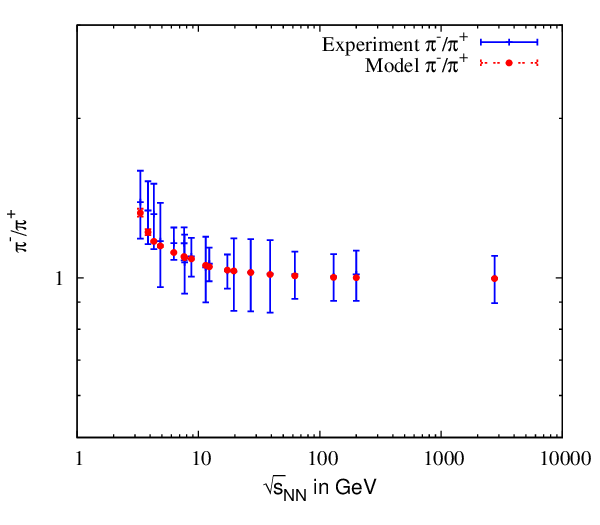}}

\caption{
\label{fg.pimpip}
Variation of $\pi^-/\pi^+$ with
$\sqrt{s}$}
\end{figure}

In Fig.~\ref{fg.pimpip} the $\pi^-$ to $\pi^+$ ratio is shown as a
function of $\sqrt{s}$. As discussed earlier, this ratio is greater than
1 for low $\sqrt{s}$ and approaches 1 at higher collision energies. Rather 
than the +ve charge of the protons the higher neutron abundance seems to 
push for the isospin asymmetry in favor of $\pi^-$. The data are well
reproduced by our analysis.

\begin{figure}[!htb]
{\includegraphics[scale=0.7]{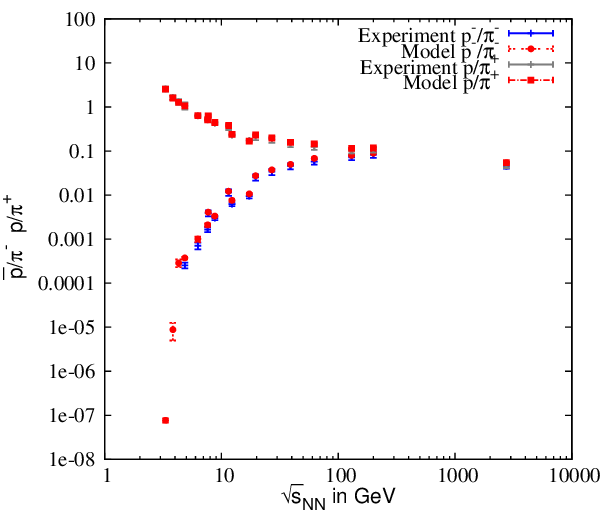}}

\caption{
\label{fg.ppi}
Variation of $p/\pi^+$ and $\bar{p}/\pi^-$ with
$\sqrt{s}$}
\end{figure}

Similarly the $p/\pi^+$ and $\bar{p}/\pi^-$ variation with $\sqrt{s}$ is
also well reproduced as shown in Fig~\ref{fg.ppi}. At lower $\sqrt{s}$,
$p > \bar{p}$ and $\pi^- > \pi^+$, while at higher $\sqrt{s}$ the
corresponding particles and antiparticles become equal. This explains
the variations shown.

\begin{figure}[!htb]
{\includegraphics[scale=0.7]{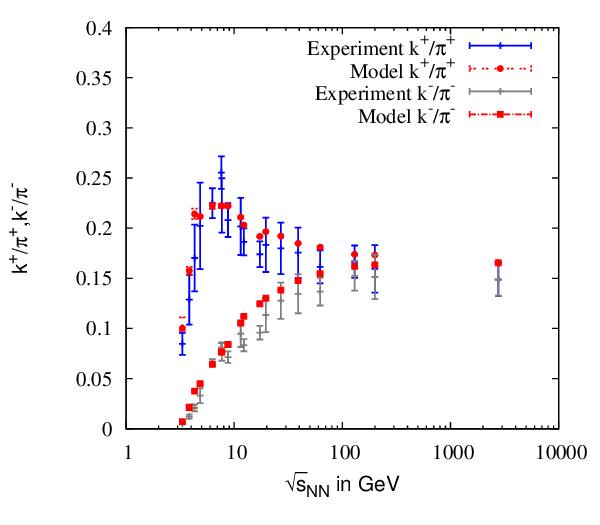}}
\caption{
\label{fg.kpia}
Variation of $k^+/\pi^+$ and $k^-/\pi^-$ with
$\sqrt{s}$}
\end{figure}

The $k/\pi$ ratio is considered to be an important observable for
strangeness enhancement in high energy collisions.  A '$horn$' in the
$k^+/\pi^+$ ratio was originally suggested as a signature of
QGP~\cite{Gazdzicki:1998vd, Gazdzicki:2003fj, Bratkovskaya:2004kv,
Koch:2005pk}. Several authors have tried to explain the behavior of
$k^+/\pi^+$ and $k^-/\pi^-$ using different approaches (see
\cite{Nayak:2010uq} and references therein). The comparison between the
experimental data for these ratios and the corresponding predictions
from our model analysis is shown in Fig.~\ref{fg.kpia}. We find that
irrespective of the underlying physical mechanism that gives rise to the
$horn$, which is beyond the scope of the HRG model, the experimental
ratios and model predictions agree quite well.

\begin{figure}[!htb]
\subfloat[]{
{\includegraphics[scale=0.7]{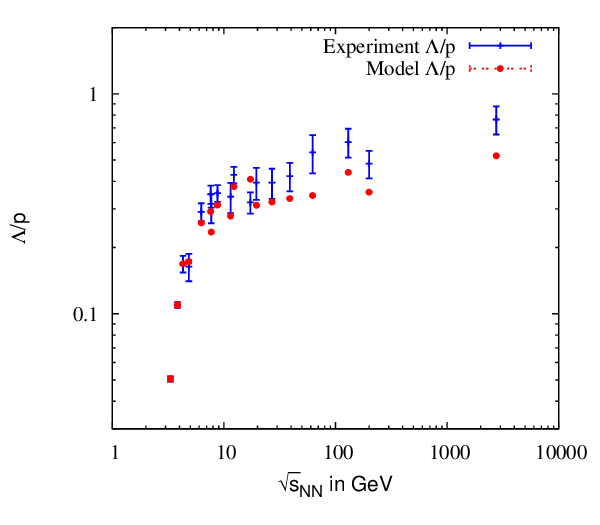}}
\label{fg.lxpratioa}
}\\
\subfloat[]{
{\includegraphics[scale=0.7]{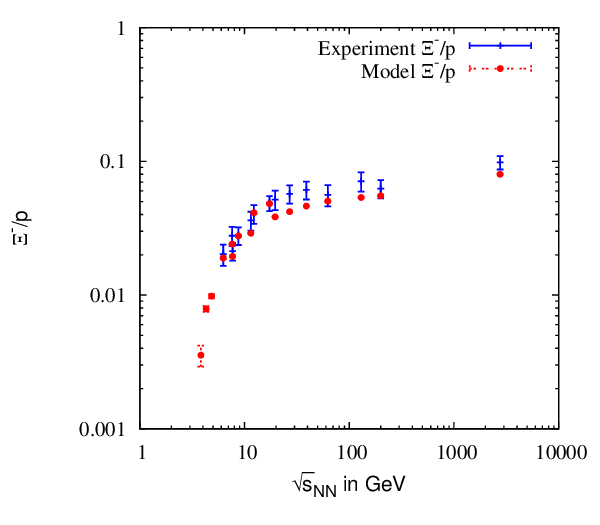}}
\label{fg.lxpratiob}
}
\caption{
\label{fg.lxpratio}
Ratios of yields of strange baryons to proton as a function of
$\sqrt{s}$.}
\end{figure}

In Fig.~\ref{fg.lxpratio} the ratios $\Lambda/p$ and $\Xi^-/p$ are
shown.  The agreement for $\Lambda/p$ is reasonable except for a slight
down-shift of the predicted results as compared to the experimental
data. Consideration of possible uncertainties in contribution from weak
decays may remove this discrepancy~\cite{Abelev:2008ab}. Such
uncertainties are not included in our analysis. The model predictions
for $\Xi^-/p$ is found to agree well with the experimental data, as
shown in Fig.~\ref{fg.lxpratiob}.

\begin{figure}[!htb]
{\includegraphics[scale=0.7]{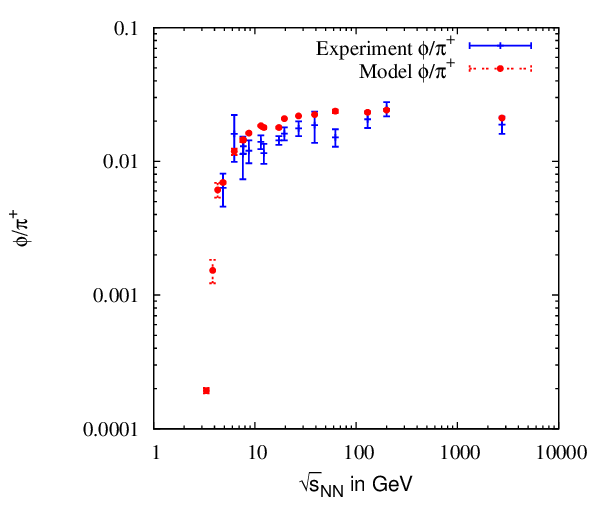}}

\caption{
\label{fg.phipi}
Variation of $\phi/\pi^+$ with
$\sqrt{s}$}
\end{figure}

Finally we present some ratios for the $\phi$, and $\Omega$ particles
whose yield data were never used in the analysis.  The $\phi/\pi^+$
ratio is shown in Fig.~\ref{fg.phipi}. Since $\phi$ has no net charge of
any kind, it is dependent only on the temperature. The predicted
$\phi/\pi^+$ plot closely resembles the temperature plot. Again
prediction from our model agrees reasonably with the experimental data.

\begin{figure}[!htb]
\subfloat[]{
{\includegraphics[scale=0.7]{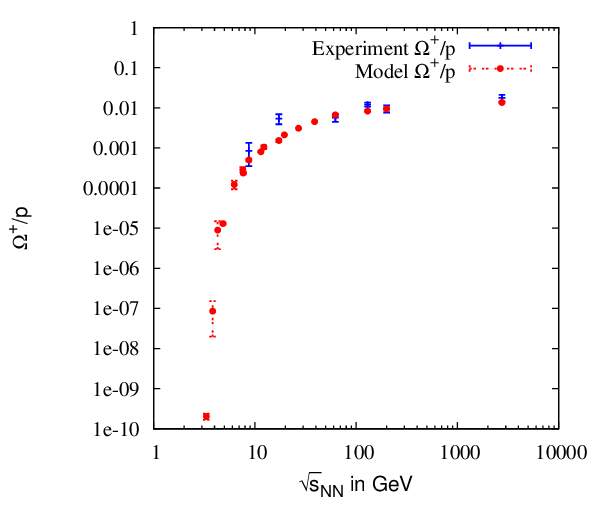}}
\label{fg.omgratioa}
}\\
\subfloat[]{
{\includegraphics[scale=0.7]{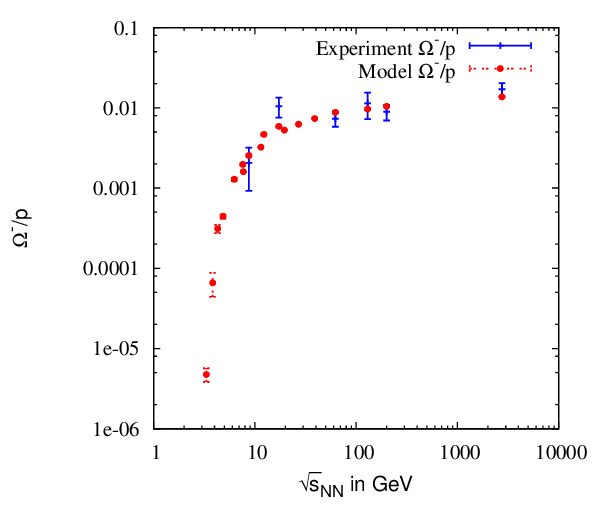}}
\label{fg.omgratiob}
}
\caption{
\label{fg.omgratio}
Ratios of yields of omega baryons to proton as a function of
$\sqrt{s}$.}
\end{figure}

The predictions for the $\Omega/p$ ratios are shown in
Fig.~\ref{fg.omgratio}. For this ratio the experimental data are
available at only a few $\sqrt{s}$. The model predictions seem to agree
quite well. 

\begin{figure}[!htb]
{\includegraphics[scale=0.7]{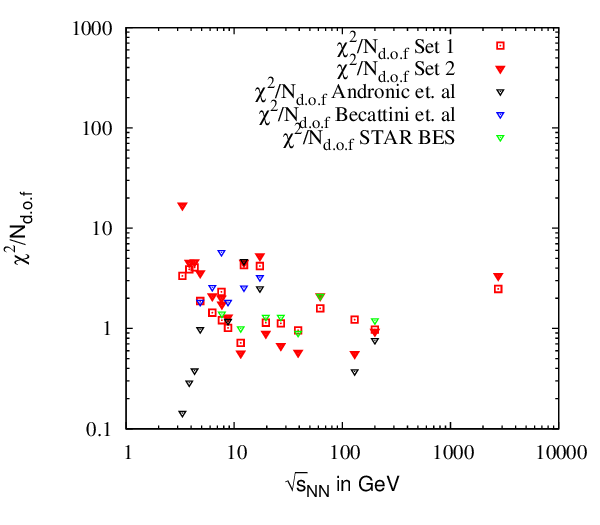}}
\caption{
\label{fg.chisquare}
Variation of $\chi^2$ }
\end{figure}

As mentioned earlier we have not done a $\chi^2$ fit. But it would be
interesting to find out the values of $\chi^2/dof$ from our estimated
freeze-out parameters. One can make various choices of {\it independent}
particle ratios.  We have used two sets $-$ one with various hadron
yields in the numerator and $\pi^+$ yield in the denominator (Set 1),
and another with different indepndent ratios $\pi^-/\pi^+$, $k^+/\pi^-$,
$k^-/k^+$, $p/k^-$, $\bar{p}/p$, $\Lambda/\bar{p}$,
$\bar{\Lambda}/\Lambda$, $\Xi^-/\bar{\Lambda}$, $\Xi^+/\Xi^-$ (Set 2).
The variation of $\chi^2/dof$ with $\sqrt{s}$ is shown in
Fig.~\ref{fg.chisquare}, along with some available results in the
literature \cite{Andronic:2005yp, Becattini:2005xt, Adamczyk:2017iwn}
We find the reduced $\chi^2$ in various approaches to be
generally in agreement with each other. 

~\\
{\it Summary and Outlook:} $-$
Here we introduced a novel formalism for parametrizing the chemical
freeze-out surface of hadrons identified from the heavy-ion collision
experiments. We have demonstrated that the conserved charges of strong
interactions may be utilized to obtain the freeze-out parameters.  We
explored the chemical freeze-out surface, in the central rapidity bins
for the most central collisions in a wide range of center of mass
energies.  The various charge ratios agree in the model and data to the
accuracy of $10^{-6}$ or better.  This strongly confirms that the
various charges themselves are in chemical equilibrium near freeze-out. 

Our formalism allows us to successfully reproduce the various
hadronic yield ratios. Even the predictions for multi-strange hadrons
like $\Omega$, whose yields were never used in the parametrization, are
predicted satisfactorily. This simple yet physically consistent approach
could be a viable alternative over the conventional $\chi^2$ scheme for
analyzing data from the upcoming heavy-ion collision experiments.

\begin{acknowledgments}
This work is funded by UGC, CSIR and DST of the Government of India. DB
thanks Sabita Das and Sandeep Chatterjee for discussion about BES data.
We thank Anirban Lahiri and Subhasis Samanta for various useful
discussions.
\end{acknowledgments}


\bibliography{ref}
\end{document}